\documentclass[]{iopart}
\usepackage{iopams,setstack}
\usepackage{graphicx}
\newcommand{\ii}{\mathrm{i}}
\newcommand{\be}{\begin{equation}}
\newcommand{\ee}{\end{equation}}

\newcommand{\ba}{\begin{array}}
\newcommand{\ea}{\end{array}}
\newcommand{\bea}{\begin{eqnarray}}
\newcommand{\eea}{\end{eqnarray}}
\newcommand{\nn}{\nonumber}
\newcommand{\R}{\mathbb{R}}

\usepackage{graphicx}
\usepackage{dcolumn}
\usepackage{bm}

\newcommand{\eqref}[1]{(\ref{#1})}

\begin{document}


\title{Three-dimensional superintegrable systems in a static electromagnetic field}

\author{A. Marchesiello$^1$, L. {\v{S}}nobl$^1$ and P. Winternitz$^2$}
\address{$^1$ Czech Technical University in Prague, Faculty of Nuclear Sciences and Physical Engineering, 
B\v rehov\'a 7, 115 19 Prague 1, Czech Republic\ead{anto.marchesiello@gmail.com, Libor.Snobl@fjfi.cvut.cz}
}
\address{$^2$
 Centre de recherches math\'ematiques and Departement de math\'ematiques et de statistique,
Universit\'e de Montr\'eal, CP 6128, Succ Centre-Ville, Montr\'eal (Qu\'ebec) H3C 3J7, Canada. 
Presently on sabbatical leave at Dipartimento di Matematica e Fisica, Universit\`a degli Studi 
Roma Tre, Via della Vasca Navale 84, 00146 Roma, Italy.\ead{wintern@crm.umontreal.ca}
}
\begin{abstract}
We consider a charged particle moving in a static electromagnetic field described by the vector potential 
$\vec A(\vec x)$ and the electrostatic potential $V(\vec x)$. We study the conditions on the structure of the integrals of motion of the first and second order in momenta, in particular how they are influenced by the gauge invariance of the problem. Next, we concentrate on the three possibilities for integrability arising from the first order integrals corresponding to three nonequivalent subalgebras of the Euclidean algebra, namely $(P_1,P_2)$, $(L_3,P_3)$ and $(L_1,L_2,L_3)$. For these cases we look for additional independent integrals of first or second order in the momenta. These would make the system superintegrable (minimally or maximally). We study their quantum spectra and classical equations of motion. In some cases nonpolynomial integrals of motion occur and ensure maximal superintegrability.\\

\end{abstract}

\section{Introduction}

The purpose of this article is to initiate a systematic study of integrable and superintegrable systems in the presence of a magnetic field in three-dimensional Euclidean space $E_3$, both in classical and quantum mechanics. 

We recall that a classical Hamiltonian system with $n$ degrees of freedom is (Liouville) integrable if it allows $n$ integrals of motion $X_1,\ldots,X_n$ including the Hamiltonian. They must be in involution, be well defined functions on phase space and be functionally independent. The system is superintegrable if further integrals $Y_a$ exist with $1\leq a\leq n-1$. They must also be well-defined functions on phase space, the set $\{X_i,Y_a\}$ must be functionally independent, however the integrals $Y_a$ need not be in involution with $X_i$ nor with each other. On the contrary, the set of all integrals generates a non-Abelian algebra under Poisson commutation. In quantum mechanics the definitions must be modified. The integrals $X_1,\ldots,X_n$ will be well-defined operators in the enveloping algebra of the Heisenberg Lie algebra, i.e. polynomials in the coordinates $x_i$ and $p_i$, or more general objects, such as convergent series in these objects. Instead of functional independence we shall require polynomial independence. In other words no nontrivial fully symmetrized polynomial (Jordan polynomial) in the integrals of motion should vanish.

The best known superintegrable systems in $E_n$ $(n\geq 2)$ are the Coulomb-Kepler system and the isotropic harmonic oscillator with their $\mathfrak{o}(n+1)$ and $\mathfrak{s}\mathfrak{u}(n)$ symmetry algebras, respectively~\cite{Fock,Bargmann,JauHi,MosSmi}. It follows from Bertrand's theorem \cite{GolPooSaf} that these are the only spherically symmetric maximally superintegrable systems in $E_3$ (and actually in $E_n$ $(n\geq 2)$). Most of the recent research on superintegrability concentrated on ``natural'' Hamiltonians with scalar potentials. For a recent review see \cite{MiPoWin}. For early systematic work on superintegrability in $E_2$ see~\cite{FriManSmoUhlWin,WinSmoUhlFr} and in $E_3$~\cite{MaSmoVaWin,Evans1,Evans2,VerEv}, see also \cite{KalMi1,KalMi2,KalMi3} and references therein.

Earlier work on integrability with magnetic fields (velocity dependent potentials) mainly concerned the two-dimensional case \cite{DoGraRaWin,BeWin,CharHuWin,McSWin,PuRos,Pucacco} with notable exceptions \cite{Zhalij,BeChaRas}.

A particle moving under the influence of a static electromagnetic field is described by the classical Hamiltonian
\begin{equation}\label{classham}
H= \frac{1}{2} (\vec p + \vec A(\vec x))^2 + V(\vec x)
\end{equation}
where $\vec p=(p_1,p_2,p_3)$ are components of the linear momentum, $\vec x=(x_1,x_2,x_3)\equiv (x,y,z)$ the spatial coordinates, the vector and electrostatic potential $\vec A(\vec x)$, $V(\vec x)$ are functions only of the coordinates $\vec x$ and we choose the units in which the mass of the particle has the numerical value $1$ and the charge of the particle is $-1$ (having an electron in mind as the prime example).

We recall that the equations of motion of the Hamiltonian~\eqref{classham} are gauge invariant, i.e. 
that they are the same for the potentials
\begin{equation}\label{classgaugeequiv}
\vec A'(\vec x)=\vec A(\vec x)+ \nabla \chi, \qquad V'(\vec x)=V(\vec x) 
\end{equation}
for any choice of the function $\chi(\vec x)$. Thus, the physically relevant quantity is the magnetic field
\begin{equation}
\vec B=  \nabla \times \vec A, \qquad \mathrm{  i.e. } \qquad  B_j=  \epsilon_{jkl} \frac{\partial A_l}{\partial x_k}
\end{equation}
rather than the vector potential $\vec A(\vec x)$ (where $\epsilon_{jkl}$ is the completely antisymmetric tensor with $\epsilon_{123}=1$).

We shall also consider the quantum Hamiltonian defined as the properly symmetrized analogue of \eqref{classham} in terms of the operators of the linear momenta $\hat P_j=- \ii \hbar \partial/{\partial x_j}$ and coordinates $\hat X_j=x_j$:
\begin{equation}\label{quantham}
\hat H= \frac{1}{2} \sum_j \left(\hat P_j \hat P_j + \hat P_j \hat A_j(\vec x)+ \hat A_j(\vec x) \hat P_j + \hat  A_j \hat A_j\right) + \hat V(\vec x).
\end{equation}
The operators $\hat A_j(\vec x)$ and $\hat V(\vec x)$ act on wavefunctions as multiplication by the functions $A_j(\vec x)$ and $V(\vec x)$, respectively.

On the quantum level, the gauge transformation~\eqref{classgaugeequiv} demonstrates itself as a unitary transformation of the Hilbert space. Namely, let us take
\begin{equation}\label{quantumgauge}
\hat U \psi(\vec x)= \exp{\left(\frac{\ii}{\hbar} \chi(\vec x)\right)} \cdot \psi(\vec x).
\end{equation}
Applying \eqref{quantumgauge} on the states and the observables we get an equivalent description of the same physical reality in terms of
\begin{equation}
 \psi \rightarrow \psi'=\hat U \psi, \qquad \hat O \rightarrow \hat O'= \hat U \hat O \hat U^{\dagger}.
\end{equation}
In particular, the following observables transform covariantly
\begin{equation}\label{covarPa}
(\hat P_j+\hat A_j)\rightarrow \hat U (\hat P_j+\hat A_j) \hat U^{\dagger} =  P_j+ \hat A'_j, \qquad  \hat V \rightarrow \hat U \hat V \hat U^{\dagger}= \hat V. 
\end{equation}
We recall that the quantum dynamics may not depend only on the magnetic field $\vec B$. In nontrivial topologies of the configuration space, e.g. when singularities are present in the potential, not all vector potentials $\vec A$ inducing the same magnetic field $\vec B$ are gauge equivalent, as the observable Aharonov-Bohm effect demonstrates~\cite{AhaBoh}.

\section{The conditions for the integrals of motion}\label{ConIntMot}
Let us consider integrals of motion which are at most second order in the momenta. Because our system is gauge invariant~\eqref{classgaugeequiv},~\eqref{covarPa} we find it convenient to express the integrals in terms of gauge covariant expressions
\begin{equation}\label{covarP}
p_j^A= p_j+A_j, \qquad \hat P_j^A= \hat P_j+\hat A_j
\end{equation}
rather than the momenta themselves. The operators \eqref{covarP} do not commute amongst each other. They satisfy
\begin{equation}\label{covarPcomms}
[\hat P_j^A,\hat P_k^A]=-\ii \hbar \epsilon_{jkl} \hat B_l, \qquad [\hat P_j^A,\hat X_k]=-\ii \hbar \delta_{jk}
\end{equation}
and analogous relations for Poisson brackets in classical mechanics.

Classically, we write a general second order integral of motion as
\begin{equation}\label{classint}
\fl X= \sum_{j=1}^{3} h_j(\vec x) p_j^A p_j^A + \sum_{j,k,l=1}^{3} \frac{1}{2} |\epsilon_{jkl}| n_j(\vec x) p_k^A p_l^A + \sum_{j=1}^{3} s_j(\vec x) p_j^A+m(\vec x).
\end{equation}
We define the Poisson bracket in the standard manner
\begin{equation}
\{a(\vec x,\vec p),b(\vec x,\vec p)\}_{P.B.}=\sum_{j=1}^{3}\left(\frac{\partial a}{\partial {x_j}} \frac{\partial b}{\partial {p_j}} - \frac{\partial b}{\partial {x_j}} \frac{\partial a}{\partial {p_j}} \right).
\end{equation}
The condition that the Poisson bracket of the integral \eqref{classint} with the Hamiltonian~\eqref{classham} vanishes
\begin{equation}
\{ H,X\}_{P.B.}=0
\end{equation}
leads to terms of order $3,2,1$ and $0$ in the momenta and respectively to the following equations:\\
 Third  order
\begin{eqnarray}
\nn \partial_x h_1  = 0, \qquad  \partial_y h_1  = -\partial_x n_3 , \qquad  \partial_z h_1  =- \partial_x n_2 ,\\ \label{HOconds} 
\partial_x h_2  =-\partial_y n_3,  \qquad \partial_y h_2  =0, \qquad  \partial_z h_2  =-\partial_y n_1 ,\\ \nn \partial_x h_3  =- \partial_z n_2 , \qquad  \partial_y h_3  =- \partial_z n_1 ,  \qquad \partial_z h_3  = 0,\\ \nn   \nabla \cdot \vec n  =0 . 
\end{eqnarray}
Second order
\begin{eqnarray}
\nn \partial_x s_1  = n_2 B_2-n_3 B_3, \\
\nn \partial_y s_2  = n_3 B_3-n_1 B_1, \\
\nn \partial_z s_3  = n_1 B_1-n_2 B_2, \\
\label{2ordcond} \partial_y s_1 + \partial_x s_2  =n_1 B_2 -n_2 B_1+2 (h_1 - h_2) B_3, \\
\nn \partial_z s_1+\partial_x s_3  = n_3 B_1-n_1 B_3+2 (h_3 - h_1) B_2, \\
\nn \partial_y s_3+\partial_z s_2  = n_2 B_3-n_3 B_2+2 (h_2 - h_3) B_1.
\end{eqnarray}
\hskip15mm It follows that we have
$$ \nabla \cdot \vec s=0. $$
First order
\begin{eqnarray}
\nn \partial_x m = 2 h_1 \partial_x V+ n_3 \partial_y V+ n_2 \partial_z V+s_3 B_2-s_2 B_3, \\
\label{1ordcond} \partial_y m = n_3 \partial_x V+2 h_2 \partial_y V+ n_1 \partial_z V+s_1 B_3-s_3 B_1, \\
\nn \partial_z m = n_2 \partial_x V+ n_1 \partial_y V+2 h_3 \partial_z V+s_2 B_1-s_1 B_2.
\end{eqnarray}
Zero order
\begin{eqnarray}\label{0ordcond}
\vec s \cdot \nabla V = 0.
\end{eqnarray}

Equations~\eqref{HOconds} are the same as for the system with vanishing magnetic field and their explicit solution is known. They imply that the highest order terms in the integral \eqref{classint} are linear combinations of products of the generators of the Euclidean group $p_1,p_2,p_3,l_1,l_2,l_3$ where $l_j=\sum_{k,l} \epsilon_{jkl} x_k p_l$ (up to a redefinition of lower order terms). When expressed explicitly in terms of the covariant expressions~\eqref{covarP}, we have
\begin{equation}\label{classintUEA}
X = \sum_{1\leq a\leq b \leq 6} \alpha_{ab} Y_a^A Y_b^A+ \sum_{j=1}^{3} s_j(\vec x) p_j^A+m(\vec x)
\end{equation}
where $Y^A=(p_1^A,p_2^A,p_3^A,l_1^A,l_2^A,l_3^A)$, 
\begin{equation}\label{covarL}
l_j^A=\sum_{k,l} \epsilon_{jkl} x_k p_l^A,
\end{equation} 
and $\alpha_{ab}\in \R$. The functions $\vec h,\vec n$ are expressed in terms of the constants $\alpha_{ab}$ as follows
\begin{eqnarray}
\fl \nn h_1 & = \alpha_{66} y^2+(-\alpha_{56} z-\alpha_{16}) y+\alpha_{55} z^2+\alpha_{15} z+\alpha_{11}, \\
\fl \nn h_2  & = \alpha_{66} x^2+(-\alpha_{46} z+\alpha_{26}) x+\alpha_{44} z^2-\alpha_{24} z+\alpha_{22}, \\
\fl \label{EuclUEA} h_3  & = \alpha_{55} x^2+(-\alpha_{45} y-\alpha_{35}) x+\alpha_{44} y^2+\alpha_{34} y+\alpha_{33}, \\
\fl \nn n_1 & = -\alpha_{56} x^2+(\alpha_{46} y+\alpha_{45} z-\alpha_{25}+\alpha_{36}) x+(-2 \alpha_{44} z+\alpha_{24}) y-\alpha_{34} z+\alpha_{23}, \\
\fl \nn n_2 & = (\alpha_{56} y-2 \alpha_{55} z-\alpha_{15}) x-\alpha_{46} y^2+(\alpha_{45} z+\alpha_{36}+\alpha_{14}) y+\alpha_{35} z+\alpha_{13}, \\
\fl \nn n_3 & = (-2 \alpha_{66} y+\alpha_{16}+\alpha_{56} z) x+(\alpha_{46} z-\alpha_{26}) y-\alpha_{45} z^2+(\alpha_{25}-\alpha_{14}) z+\alpha_{12}.
\end{eqnarray}
Since we have the relation
\begin{equation}\label{pl0}
\vec p \cdot \vec l=0 \end{equation}
one of the parameters $\alpha_{14},\alpha_{25},\alpha_{36}$ is irrelevant and can be set equal to any chosen value. However that explicitly breaks the symmetry of our equations; thus we prefer to keep all parameters $\alpha_{14},\alpha_{25},\alpha_{36}$ in our formulas and we use~\eqref{pl0} as a constraint.

The conditions~\eqref{2ordcond} allow us to express all the second order derivatives of $\vec s$ explicitly in terms of $\vec n$, $\vec h$, $\vec B$ and their derivatives
\begin{eqnarray}
\fl \nn \partial^2_{xx} s_1  = & \partial_x n_2 B_2+ n_2 \partial_x B_2-\partial_x n_3 B_3-n_3 \partial_x B_3,\\
\fl \nn \partial^2_{xy} s_1  = &\partial_y n_2 B_2+n_2 \partial_y B_2-\partial_y n_3 B_3- n_3 \partial_y B_3,\\
\fl \nn \partial^2_{xz} s_1  = & \partial_z n_2 B_2+ n_2 \partial_z B_2-\partial_z n_3 B_3-n_3 \partial_z B_3,\\ 
\fl \nn \partial^2_{yy} s_1  = &  n_1 \partial_x B_1-3 \partial_x n_3 B_3-n_3 \partial_x B_3+\partial_x n_1 B_1+2 h_1 \partial_y B_3-\partial_y n_2 B_1 \\ \nn 
\fl & -n_2 \partial_y B_1+n_1 \partial_y B_2-2 h_2 \partial_y B_3+\partial_y n_1 B_2,\\ 
\fl \nn \partial^2_{yz} s_1  = & -\frac{3}{2} \partial_z n_2 B_1-\frac{1}{2} n_2 \partial_z B_1+\frac{3}{2} \partial_y n_3 B_1+\frac{1}{2} n_3 \partial_y B_1-\frac{3}{2} \partial_x n_2 B_3 \\ 
\fl \nn & -\frac{1}{2} n_2 \partial_x B_3+h_1 \partial_z B_3+\frac{1}{2} \partial_y n_1 B_3-h_2 \partial_z B_3-\frac{1}{2} \partial_z n_1 B_2+\frac{1}{2} n_1 \partial_z B_2 \\ 
\fl \nn & -\frac{1}{2} n_1 \partial_y B_3+\frac{3}{2} \partial_x n_3 B_2-h_1 \partial_y B_2+h_3 \partial_y B_2+\frac{1}{2} n_3 \partial_x B_2-h_2 \partial_x B_1+h_3 \partial_x B_1,\\ 
\fl \nn \partial^2_{zz} s_1 = & 3 \partial_x n_2 B_2+ n_2 \partial_x B_2-\partial_x n_1 B_1-n_1 \partial_x B_1-2 h_1 \partial_z B_2-\partial_z n_1 B_3 \\ 
\fl \nn & - n_1 \partial_z B_3+ n_3 \partial_z B_1+2 h_3 \partial_z B_2+\partial_z n_3 B_1,
\\ 
\fl  \nn \partial^2_{xx} s_2 = & -\partial_y n_2 B_2-n_2 \partial_y B_2+3 \partial_y n_3 B_3+n_3 \partial_y B_3+2 h_1 \partial_x B_3-\partial_x n_2 B_1 
\\ 
\fl \nn & -n_2 \partial_x B_1+ n_1 \partial_x B_2-2 h_2 \partial_x B_3+\partial_x n_1 B_2, \\
\fl  \nn \partial^2_{xy} s_2 = & \partial_x n_3 B_3+n_3 \partial_x B_3-\partial_x n_1 B_1-n_1 \partial_x B_1, \\
\fl \nn \partial^2_{xz} s_2 = & \frac{1}{2} \partial_z n_2 B_1-\frac{1}{2} n_2 \partial_z B_1-\frac{3}{2} \partial_y n_3 B_1-\frac{1}{2} n_3 \partial_y B_1-\frac{1}{2} \partial_x n_2 B_3 +\frac{1}{2} n_2 \partial_x B_3 
\\ 
\fl \nn &
+h_1 \partial_z B_3  +\frac{3}{2} \partial_y n_1 B_3-h_2 \partial_z B_3+\frac{3}{2} \partial_z n_1 B_2+\frac{1}{2} n_1 \partial_z B_2+\frac{1}{2} n_1 \partial_y B_3
\\ 
\fl \nn & -\frac{3}{2} \partial_x n_3 B_2+h_1 \partial_y B_2 -h_3 \partial_y B_2-\frac{1}{2} n_3 \partial_x B_2+h_2 \partial_x B_1-h_3 \partial_x B_1, \\
\fl \nn \partial^2_{yy} s_2 = & \partial_y n_3 B_3+n_3 \partial_y B_3-\partial_y n_1 B_1-n_1 \partial_y B_1, \\
\fl \label{2ndderss} \partial^2_{yz} s_2 = & \partial_z n_3 B_3+n_3 \partial_z B_3-\partial_z n_1 B_1-n_1 \partial_z B_1, \\
\fl \nn \partial^2_{zz} s_2 = & -n_1 \partial_y B_1+\partial_y n_2 B_2+n_2 \partial_y B_2-3 \partial_y n_1 B_1+2 h_2 \partial_z B_1-\partial_z n_3 B_2 
\\ \fl  \nn & -n_3 \partial_z B_2 -2 h_3 \partial_z B_1  +\partial_z n_2 B_3+n_2 \partial_z B_3, \\
\fl \nn \partial^2_{xx} s_3 = & n_3 \partial_z B_3-3 \partial_z n_2 B_2-n_2 \partial_z B_2+\partial_z n_3 B_3-2 h_1 \partial_x B_2-\partial_x n_1 B_3 \\ \nn 
\fl & -n_1 \partial_x B_3 +n_3 \partial_x B_1  +2 h_3 \partial_x B_2+\partial_x n_3 B_1, \\
\fl \nn \partial^2_{xy} s_3  = & \frac{3}{2} \partial_z n_2 B_1
+\frac{1}{2} n_2 \partial_z B_1-\frac{1}{2} \partial_y n_3 B_1+\frac{1}{2} n_3 \partial_y B_1+\frac{3}{2} \partial_x n_2 B_3 +\frac{1}{2} n_2 \partial_x B_3
\\ \fl  \nn & -h_1 \partial_z B_3-\frac{3}{2} \partial_y n_1 B_3+h_2 \partial_z B_3-\frac{3}{2} \partial_z n_1 B_2 -\frac{1}{2} n_1 \partial_z B_2 -\frac{1}{2} n_1 \partial_y B_3 
\\ \fl  \nn & +\frac{1}{2} \partial_x n_3 B_2-h_1 \partial_y B_2+h_3 \partial_y B_2-\frac{1}{2} n_3 \partial_x B_2+h_2 \partial_x B_1-h_3 \partial_x B_1,\\
\fl \nn \partial^2_{xz} s_3 = & -\partial_x n_2 B_2-n_2 \partial_x B_2+\partial_x n_1 B_1+n_1 \partial_x B_1, \\
\fl \nn \partial^2_{yy} s_3 = & -\partial_z n_3 B_3-n_3 \partial_z B_3+3 \partial_z n_1 B_1+n_1 \partial_z B_1+2 h_2 \partial_y B_1-\partial_y n_3 B_2
\\ \fl  \nn & -n_3 \partial_y B_2-2 h_3 \partial_y B_1+\partial_y n_2 B_3+n_2 \partial_y B_3, \\
\nn \fl  \partial^2_{yz} s_3 = & -\partial_y n_2 B_2-n_2 \partial_y B_2+\partial_y n_1 B_1+n_1 \partial_y B_1, \\
\nn \fl  \partial^2_{zz} s_3 = & -\partial_z n_2 B_2-n_2 \partial_z B_2+\partial_z n_1 B_1+n_1 \partial_z B_1.
\end{eqnarray}
Taking various first and second order derivatives of these expressions and comparing them, e.g.
$$ \partial_y (\partial^2_{xx} s_j)= \partial_x (\partial^2_{xy} s_j),$$
we arrive at compatibility conditions for $\vec h,\vec n$ and $\vec B$. Altogether there are 6 independent second order PDEs relating them which can be solved e.g. for 
$$\partial_{yy} B_1,\quad \partial_{zz} B_1,\quad \partial_{xx} B_2,\quad \partial_{zz} B_2,\quad \partial_{xx} B_3,\quad \partial_{yy} B_3$$
but they are too cumbersome to reproduce here. \bigskip

In the quantum case we have to consider a properly symmetrized analogue of~\eqref{classint}.
We choose the following convention
\begin{equation}\label{quantint}
\fl \hat X= \sum_{j=1}^{3} \{h_j(\vec x), \hat P_j^A \hat P_j^A\} + \sum_{j,k,l=1}^{3} \frac{|\epsilon_{jkl}|}{2}  \{n_j(\vec x), \hat P_k^A \hat P_l^A \} + \sum_{j=1}^{3} \{s_j(\vec x), \hat P_j^A\}+m(\vec x)
\end{equation}
where $\{\, , \,\}$ denotes the symmetrization 
$$
\{\hat a,\hat b\}= \frac{1}{2} \left( \hat a  \hat b + \hat b \hat a\right)
$$
and $h_j(\vec x),n_j(\vec x),s_j(\vec x),m(\vec x)$ are to be interpreted as the corresponding operators of multiplication by the given function. All possible choices of symmetrization are equivalent up to redefinition of the lower order terms.

Imposing the condition that the integral of motion in the form~\eqref{quantint} commutes with the Hamiltonian~\eqref{quantham} we obtain a similar set of conditions as above. In particular, the conditions~\eqref{HOconds} remain the same, i.e. their solution has the same form~\eqref{EuclUEA}. The conditions~\eqref{2ordcond} get apparent quantum corrections
\begin{eqnarray}
\nn \partial_x s_1 = n_2 B_2-n_3 B_3, \\
\nn \partial_y s_2 = n_3 B_3-n_1 B_1, \\
\nn \partial_z s_3 = n_1 B_1-n_2 B_2, \\
\label{2ordcondkvant} 
\partial_y s_1 + \partial_x s_2 =n_1 B_2 -n_2 B_1+2 (h_1 - h_2) B_3+\ii \hbar \partial^2_{xx} n_3, \\
\nn \partial_z s_1+\partial_x s_3 = n_3 B_1-n_1 B_3+2 (h_3 - h_1) B_2+\ii \hbar \partial^2_{xx} n_2, \\
\nn \partial_y s_3+\partial_z s_2 = n_2 B_3-n_3 B_2+2 (h_2 - h_3) B_1+\ii \hbar \partial^2_{yy} n_1. 
\end{eqnarray}
Due to the explicit solution~\eqref{EuclUEA} we have 
$$\partial^2_{xx} n_3=0,\qquad \partial^2_{xx} n_2=0,\qquad \partial^2_{yy} n_1=0,$$
i.e. these apparent quantum corrections vanish and equations~\eqref{2ordcondkvant} and~\eqref{2ordcond} are the same.

Similarly, in the quantum version of equations~\eqref{1ordcond} we obtain some extra terms but they vanish once the solution~\eqref{EuclUEA} of equations~\eqref{HOconds} is substituted in, i.e. also the conditions~\eqref{1ordcond} remain in the quantum case. 

The situation is however different for the last equation~\eqref{0ordcond} which does obtain $\hbar^2$--proportional corrections. Using equations \eqref{HOconds}, \eqref{2ordcond}, \eqref{1ordcond} and their differential consequences we find
\begin{eqnarray}\label{0ordcondquant}
 \vec s \cdot \nabla V
+ \frac{\hbar^2}{4}
\left(
\partial_z n_1 \partial_z B_1- \partial_y n_1 \partial_y B_1+  \partial_x n_2 \partial_x B_2-\partial_z n_2 \partial_z B_2 +\right. \\ \nn \left.
 +\partial_y n_3 \partial_y B_3-\partial_x n_3 \partial_x B_3  + \partial_x n_1 \partial_y B_2- \partial_y n_2 \partial_x B_1
\right)  =0.
\end{eqnarray}
Notice that the last line seems to violate the obvious symmetry of our problem under Euclidean transformations. However, this is not the case in view of the identity
\begin{equation}
\fl \partial_x n_1 \partial_y B_2- \partial_y n_2 \partial_x B_1= \partial_y n_2 \partial_z B_3- \partial_z n_3 \partial_y B_2=\partial_z n_3 \partial_x B_1- \partial_x n_1 \partial_z B_3
\end{equation}
which is a consequence of
$$ \nabla \cdot \vec B=0, \qquad \nabla \cdot \vec n=0.\medskip$$

Notice that for the special case of first order integrals the conditions \eqref{HOconds}--\eqref{0ordcond} become significantly simpler. Namely, the conditions \eqref{HOconds} do not arise. The right hand sides of the conditions \eqref{2ordcond} vanish and thus equations \eqref{2ordcond} imply that the first order term $\sum_{j=1}^{3} s_j(\vec x) p_j^A$ in the integral is a constant linear combination of the covariant linear and angular momenta $p_1^A,p_2^A,p_3^A,l_1^A,l_2^A,l_3^A$ (see~\eqref{covarP}, \eqref{covarL}). The conditions~\eqref{1ordcond} simplify to
\begin{equation}
\fl \partial_x m = s_3 B_2-s_2 B_3, \qquad \partial_y m = s_1 B_3-s_3 B_1, \qquad  \partial_z m = s_2 B_1-s_1 B_2
\end{equation}
and imply first order compatibility conditions relating $\vec B$ and $\vec s$
\begin{eqnarray}
\fl \nn \partial_y s_3  B_2+s_3 \partial_y B_2 -\partial_y s_2  B_3-s_2 \partial_y B_3 -\partial_x s_1  B_3-s_1 \partial_x B_3 +\partial_x s_3  B_1+s_3 \partial_x B_1 =0, \\
\fl \partial_z s_1  B_3+s_1 \partial_z B_3 -\partial_z s_3  B_1-s_3 \partial_z B_1 -\partial_y s_2  B_1-s_2 \partial_y B_1 +\partial_y s_1  B_2+s_1 \partial_y B_2 =0, \\
\fl \nn \partial_z s_3  B_2+s_3 \partial_z B_2 -\partial_z s_2  B_3-s_2 \partial_z B_3 -\partial_x s_2  B_1-s_2 \partial_x B_1 +\partial_x s_1  B_2+s_1 \partial_x B_2=0 .
\end{eqnarray}
The condition~\eqref{0ordcond} remains the same as for the second order integral. In this case it gets no quantum correction, i.e. \eqref{0ordcond} and \eqref{0ordcondquant} now coincide.\bigskip

Let us now turn our attention to the situation where the Hamiltonian~\eqref{classham} or~\eqref{quantham} is integrable in the Liouville sense, with at most quadratic integrals. That means that in addition to the Hamiltonian itself there must be at least two independent integrals of motion of the form~\eqref{classint} or \eqref{quantint} which commute in the sense of the Poisson bracket or Lie commutator, respectively. Independence is to be understood as functional independence in the classical situation and in the sense that no nontrivial fully symmetrized polynomial in the given operators vanishes in the quantum case.

Since the highest order conditions~\eqref{HOconds} are the same whether or not there is a magnetic field present, the first step of the analysis can be performed as in \cite{MaSmoVaWin}, leading to 11 nonequivalent possibilities for the functions $\vec h,\vec n$. Next one should look into equations~\eqref{2ordcond}, \eqref{1ordcond}, \eqref{0ordcond} (resp.~\eqref{0ordcondquant}) and their consequences to determine the nonequivalent possible choices of the magnetic field $\vec B(\vec x)$ and the electrostatic potential $V(\vec x)$. This is the approach used by A. Zhalij in \cite{Zhalij} for the special case of the integrals $X_1=P_1^2+\ldots, X_2=P_2^2+\ldots$

We shall follow a different route here. Keeping in mind that our main goal is to arrive at examples of superintegrable systems with nonvanishing magnetic field we shall assume that the integrability arises in the simplest way possible. Namely, we assume that there are at least two independent first order integrals for our Hamiltonian.

Assuming that we have
\begin{equation}
X_1=\gamma^i_1 l_i^A+\beta^i_1 p_i^A+m_1(\vec x), \qquad X_2=\gamma^i_2 l_i^A+\beta^i_2 p_i^A+m_2(\vec x)
\end{equation}
(or its quantum analogue) we may use the Euclidean transformations to simplify $X_1,X_2$. Another allowed transformation is replacing $X_1,X_2$ by an  arbitrary regular linear combination,
$$ X_1\rightarrow \tilde X_1=\kappa_1 X_1+\kappa_2 X_2,\qquad X_2\rightarrow \tilde X_2=\lambda_1 X_1+\lambda_2 X_2,\qquad \det\left( \begin{array}{cc} \kappa_1 & \kappa_2 \\ \lambda_1 & \lambda_2 \end{array}\right)\neq 0.$$
For convenience, we redefine the yet unknown functions $m_1(\vec x),m_2(\vec x)$ as needed without renaming them.

We arrive at the following possibilities
\begin{itemize}
\item If we have $\vec \gamma_1=\vec \gamma_2= 0$ then we can set $X_1$ and $X_2$ by rotation and linear combination to
\begin{equation}\label{p1p2}
X_1=p_1^A+m_1(\vec x),\qquad X_2=p_2^A+m_2(\vec x).
\end{equation} 
\item If $(\vec \gamma_1,\vec \gamma_2)\neq (\vec 0,\vec 0)$ we can transform e.g. $X_1$ by rotation and translation  into $X_1=l_3^A+\beta p_3^A+m_1(\vec x)$. 
\begin{itemize}
\item Assuming that the integrability arises directly at the first order, i.e. that $\{ X_1, X_2 \}_{P.B.}=0$, we arrive at a single possibility
\begin{equation}\label{l3p3}
X_1=l_3^A+m_1(\vec x),\qquad X_2=p_3^A+m_2(\vec x).  
\end{equation}
\item However, there is another option - to allow $X_1$ and $X_2$ to be not in involution and expect the second commuting integral to arise via Poisson brackets and polynomial combinations of $X_1,X_2$. Thus we may up to rotation and linear combination take
\begin{equation}\label{X1x2noncomm}
\fl X_1=l_3^A+\beta p_3^A+m_1(\vec x), \qquad X_2=\sigma l_1^A+\beta_2^i p_i^A+m_2(\vec x), \qquad \sigma=0,1.
\end{equation}
In order to have nontrivial dynamics, i.e. nontrivial electric and/or magnetic field, we cannot have the full Euclidean algebra represented in terms of the integrals of motion. Thus we must require that the algebra generated by the highest order terms $l_3+\beta p_3$ and $\sigma l_1+\beta_2^i p_i$ in~\eqref{X1x2noncomm} via Poisson brackets closes as a proper subalgebra of the Euclidean algebra. We have the following options:
\begin{enumerate} 
\item The algebra isomorphic to $\mathfrak{s}\mathfrak{u}(2)$
\begin{eqnarray}
\nn X_1   =l_3^A+m_1(\vec x),\quad X_2=l_1^A+m_2(\vec x),\\
\label{l1l2l3} X_3  =\{X_1,X_2\}_{P.B.}=l_2^A+m_3(\vec x).
\end{eqnarray}
\item The algebra isomorphic to the Euclidean algebra $\mathfrak{e}_2$
\begin{eqnarray*}
X_1  =l_3^A+p_3^A+m_1(\vec x),\qquad X_2=p_1^A+m_2(\vec x),\\
X_3 =\{X_1,X_2\}_{P.B.}=p_2^A+m_3(\vec x).
\end{eqnarray*}
This case is, however, already included in~\eqref{p1p2} as a special subcase.
\end{enumerate}
\end{itemize}
\end{itemize}

\section{Superintegrability for the integrable system with integrals $P_1,P_2$}

Let us start our detailed investigation by considering the case of the integrals~\eqref{p1p2}
$$
X_1=p_1^A+m_1(\vec x),\qquad X_2=p_2^A+m_2(\vec x).
$$
The condition that $X_1$ and $X_2$ are in involution is equivalent to
\begin{equation}\label{p1p2b3}
\partial_y m_1-\partial_x m_2=B_3.
\end{equation}
Equations~\eqref{1ordcond} reduce to
\begin{eqnarray}
\nn \partial_x m_1  = 0,\qquad   \partial_y m_1  = B_3, \qquad  \partial_z m_1  = -B_2,\\
\partial_x m_2  = -B_3, \qquad  \partial_y m_2  = 0, \qquad  \partial_z m_2  = B_1
\end{eqnarray}
and together with~\eqref{p1p2b3} imply that
$$B_3(\vec x)=0, \qquad B_j(\vec x)=B_j(z), \qquad  m_j(\vec x)=m_j(z), \qquad j=1,2.$$
Writing the components of the magnetic field as
\begin{equation}\label{p1p2B}
B_j(\vec x)=F'_j(z), \qquad j=1,2,
\end{equation}
we have 
\begin{equation}
m_j(\vec x)=F_j(z), \qquad j=1,2.
\end{equation}
We choose a suitable vector potential in the form (satisfying the Coulomb gauge condition $\nabla \vec A=0$)
\begin{equation}
\vec A(\vec x)=\left(F_2(z),-F_1(z), 0\right)
\end{equation}
and from the conditions~\eqref{0ordcond} we find that 
\begin{equation}
V(\vec x)=V(z).
\end{equation}
Plugging all the information obtained about functions $\vec A,\vec B, m_j$ into the assumed form of the integrals~\eqref{p1p2} we find a very simple solution (unique up to the choice of gauge)
\begin{equation}
X_1=p_1,\qquad X_2=p_2.
\end{equation}
The same result arises also in the quantum case, via essentially the same arguments.\bigskip

Let us now assume that our system~\eqref{classham} with the potentials
$$ \vec A(\vec x)=\left(F_2(z),-F_1(z), 0\right), \qquad V(\vec x)=V(z)$$
is superintegrable, i.e. that an additional independent integral of motion exists. For simplicity, let us assume that it is of first order in momenta. Up to addition of $X_1$ and $X_2$ we have
\begin{equation}\label{p1p2x3}
X_3=\gamma^i l_i^A+\beta p_3^A+m_3(\vec x).
\end{equation}
We consider equations~\eqref{1ordcond} and their compatibility conditions which take the form
\begin{eqnarray}
\nn  \gamma_2 x F''_1-\gamma_1 y F''_1-\beta F''_1- \gamma_3 F'_2 =0, \qquad  \gamma_1 F'_2-\gamma_2 F'_1  =0, \\
 -\gamma_2 x F''_2+\gamma_1 y F''_2-\gamma_3 F'_1 +\beta F''_2  =0.  
\end{eqnarray}
We arrive at two distinct possibilities:
\begin{itemize}
\item If $\gamma_1^2+\gamma_2^2\neq 0$ then $F''_1=F''_2=0$, i.e. the magnetic field~\eqref{p1p2B} is constant. This case has already been well studied in the literature, see e.g.~\cite{LanLif,Newton}. Solving equations~\eqref{1ordcond} and \eqref{0ordcond} we find that the electrostatic potential is constant too, i.e. we have a motion in constant magnetic field and no electric field. This system is superintegrable and exactly solvable as follows.
Without loss of generality we can rotate the coordinate system so that we have
\begin{equation}\label{constBpots}
\vec B(\vec x)=(B,0,0), \qquad \vec A(\vec x)=(0,-B \, z, 0), \qquad V(\vec x)=0.
\end{equation}
Four independent first order integrals exist in this case. We write them down in the classical situation, quantum mechanically they are the same expressions in terms of operators. They read
\begin{equation}\label{constBints}
\fl X_1=p_1, \qquad X_2=p_2, \qquad X_3=p_3-B y, \qquad X_4=l_1+\frac{B}{2}(z^2-y^2).
\end{equation}
The Hamiltonian can be expressed in terms of $X_1,\ldots,X_4$ as
\begin{equation}
\label{constBHam} \fl H=\frac{1}{2} (p_1^2 +p_2^2 +p_3^2) -B z p_2 + \frac{B^2}{2} z^2  =\frac{1}{2}\left(X_1^2+X_2^2+X_3^2\right)+B X_4.
\end{equation}
The classical equations of motion 
\begin{eqnarray}
\dot x = p_1, \qquad & \dot y  = p_2-B z, \qquad & \dot z  = p_3,\\
\nn \dot  p_1 = 0, \qquad & \dot  p_2  = 0,\qquad & \dot  p_3  = B (p_2-Bz)
\end{eqnarray}
are solved explicitly as
\begin{eqnarray}
\fl \nn x(t) = x_0+p_1^0 t,\\ 
\label{constBsoln} \fl y(t) = y_0-\frac{p_3^0}{B}+\cos(B t) \frac{p_3^0}{B}-\sin(B t)\left(z_0-\frac{p_2^0}{B}\right),\\
\fl \nn  z(t) = \frac{p_2^0}{B} + \sin(B t)\frac{p_3^0}{B} +\cos(B t)\left( z_0-\frac{p_2^0}{B}\right), \\
\fl \nn p_1(t) = p_1^0, \qquad p_2(t) = p_2^0 , \qquad p_3(t) = \cos(B t) p_3^0+\sin(B t) (p_2^0-B  z_0)
\end{eqnarray}
where $(x_0,y_0,z_0)$ are the initial coordinates and $(p_1^0,p_2^0,p_3^0)$ the initial momenta.   From~\eqref{constBsoln} we find that the trajectory is a helix with axis parallel to the $x$--axis and the integrals $X_2,X_3,X_4$ determine the diameter and position of the enveloping cylinder in the $yz$--plane
\begin{equation}
\left(y+\frac{X_3}{B}\right)^2+\left(z-\frac{X_2}{B}\right)^2=\frac{1}{B^2} \left(X_2^2+X_3^2+2 B X_4\right).
\end{equation}
When $p_1=0$ the helix collapses into a circle in the plane $x=x_0$. Thus the problem reduces to the two-dimensional one. Let us now restrict to the case $p_1\neq 0$.

By inspection of the solution of the equations of motion one finds that this system is maximally superintegrable with, however, the fifth independent integral not polynomial in momenta. It reads
\begin{equation}\label{constB5int}
X_5=(B z-p_2)\cos\left(\frac{B x}{p_1}\right)- p_3 \sin\left(\frac{B  x}{p_1}\right).
\end{equation}
How to interpret this integral in the quantum case is not too clear. However, if we restrict ourselves to the subspace in the Hilbert space defined by the constraint
\begin{equation}
\hat P_1 \psi(\vec x)= k_1 \psi(\vec x), \qquad k_1\neq 0
\end{equation}
we can expand $\hat X_5$ into a convergent Taylor series in $\frac{B x}{k_1}$. We can then interpret the quantum integral of motion $\hat X_5$ as an operator in the ``extended enveloping algebra'' of the Heisenberg algebra. 

Alternatively, for the classical Hamiltonian we can perform a canonical transformation in the $x,p_1$ plane
\begin{equation}
\tilde p_1=\frac{p_1^2}{2}, \qquad \tilde x=\frac{x}{p_1}, \qquad p_1>0
\end{equation}
which transforms the integrals into
\begin{eqnarray}
\nn \tilde H  =\tilde p_1+\frac{1}{2}\left( (p_2-B z)^2+p_3^2\right), \qquad \tilde X_1  =\tilde p_1,\qquad  \tilde X_2  = p_2,\\
\tilde X_3= p_3-B y, \qquad  \tilde X_4  =l_1+\frac{B}{2} \left( z^2-y^2 \right), \\
\nn  \tilde X_5 = (B z-p_2)\cos\left(B \tilde x\right)- p_3 \sin\left(B \tilde x\right).
\end{eqnarray}
Thus $\tilde X_5$ becomes a first order polynomial in the momenta. The price is that $\tilde H$ no longer  has the ``natural'' form~\eqref{constBHam}. \medskip

The integrals $X_1,\ldots,X_5$ give rise to a Lie algebra of integrals of motion in the following manner.
We define additional functionally dependent integrals
\begin{equation}
\fl X_6= \{X_4,X_5 \}_{P.B.}=(p_2-B z) \sin\left(\frac{B x}{p_1}\right)-p_3 \cos\left(\frac{B x}{p_1}\right), \qquad X_7=1
\end{equation}
and redefine the first integral to be
\begin{equation}
\tilde X_1=\frac{X_1^2}{2}=\frac{p_1^2}{2}.
\end{equation}
The Poisson brackets now form a 7--dimensional Lie subalgebra of integrals of motion in the algebra of observables on our system, with the Lie brackets
\begin{equation}\label{tabcomms}
\fl \begin{array}{|c|c|c|c|c|c|c|c|} \hline 
\{  \, , \, \}_{P.B.}  & \tilde X_1& X_2& X_3& X_4& X_5& X_6& X_7 \\ \hline
 \tilde X_1& 0& 0& 0& 0& - B X_6&   B X_5 & 0\\ \hline
 X_2& 0& 0& B X_7& -X_3& 0& 0& 0\\ \hline
 X_3& 0& -B X_7& 0& X_2& 0& 0& 0\\ \hline
 X_4& 0& X_3& -X_2& 0& X_6& -X_5& 0\\ \hline
 X_5& B X_6& 0& 0& -X_6& 0& -B X_7& 0\\ \hline
 X_6& - B X_5& 0& 0& X_5& B X_7& 0& 0\\ \hline
 X_7& 0& 0& 0& 0& 0& 0& 0 \\ \hline
\end{array}
\end{equation}
This algebra is solvable with 5--dimensional nilradical spanned by $X_2,X_3,X_5,X_6,X_7$. The nilradical is isomorphic to the Heisenberg algebra in two spatial dimensions ($\mathfrak{n}_{5,3}$ in the notation of~\cite{SnoWi}). The element $X_7$ spans the center. Its Casimir invariants are the central element $X_7$ and two second order invariants $2 \tilde X_1 X_7+X_5^2+X_6^2$, $2(B X_4+\tilde X_1) X_7+X_2^2+X_3^2$ which both reduce to the Hamiltonian~\eqref{constBHam} once the explicit form of the integrals is inserted into them.

We observe that in terms of the original integral $X_1$ instead of its square $\tilde X_1$ we obtain an infinite-dimensional loop algebra.

As it is well-known in the physics literature (see e.g.~\cite{Newton}, pg. 220), the Schr\"odinger equation for the Hamiltonian~\eqref{constBHam} separates in Cartesian coordinates as follows
\begin{eqnarray}
\nn \psi(\vec x) = f(z) \exp\left(\frac{\ii}{\hbar} k_1 x\right) \exp\left(\frac{\ii}{\hbar} k_2 y\right),\\
\label{reducedconstB} \hbar^2 \ddot f(z) = \left( (B z- k_2)^2 +k_1^2-2 E \right) f(z),\\
\nn X_{1} \psi(\vec x)=k_{1} \psi(\vec x), \qquad X_{2} \psi(\vec x)=k_{2} \psi(\vec x). 
\end{eqnarray}
The reduced Schr\"odinger equation \eqref{reducedconstB} is the stationary Schr\"odinger equation for the 1--dimensional harmonic oscillator with the energy $E-\frac{k_1^2}{2}$, frequency $\omega=B$ and the center of the force at $z=k_2/B$. Thus the spectrum of the Hamiltonian~\eqref{constBHam} is continuous due to the arbitrary momentum $k_1$ and reads
\begin{equation}
\label{constBspectrum} E=\frac{k_1^2}{2} + \hbar B \left(n+\frac{1}{2}\right), \qquad n\in\mathbb{N}_{0}, \; k_1\in \mathbb{R}.
\end{equation}
The eigenvectors are expressed in terms of Hermite polynomials
\begin{equation}
\label{constBeigenvectors} \fl \psi_{n,k_1,k_2}(\vec x) = H_n\left(\sqrt{\frac{B}{\hbar}}\left(z-\frac{k_2}{B}\right)\right) \exp\left(\frac{\ii}{\hbar} \left(k_1 x+ k_2 y\right)\right) \exp\left( -\frac{B}{2 \hbar} \left(z-\frac{k_2}{B}\right)^2 \right).
\end{equation}
It was conjectured in~\cite{TTW} that all maximally superintegrable quantum systems in Euclidean spaces are exactly solvable. Equations~\eqref{constBspectrum}, \eqref{constBeigenvectors} confirm this conjecture for a particle in a constant magnetic field. This is true even though one of the independent integrals $X_5$ is not polynomial in its natural setting.

\item If $\gamma_1=\gamma_2=0$ we have to distinguish further subcases. If $\gamma_3=0$ the system again reduces to a constant magnetic field and vanishing electric field. If $\gamma_3\neq 0$ and $\beta=0$ the magnetic field must vanish and we are in a situation without vector potential, which is not of interest here (and already well studied, see e.g. \cite{MaSmoVaWin,Evans1,Evans2,VerEv}). However, if $\gamma_3\neq 0, \, \beta\neq 0$ (and without loss of generality we can assume $\gamma_3= 1$) we obtain a nontrivial solution for $\vec A$ and  $V$
\begin{eqnarray}
 \nn \vec A(\vec x) =\left(-A \cos\left(\frac{z+\phi_0}{\beta} \right),-A \sin\left(\frac{z+\phi_0}{\beta} \right),0\right), \\
 \vec B(\vec x)=\left(\frac{A}{\beta} \cos\left( \frac{z+\phi_0}{\beta} \right), \frac{A}{\beta} \sin\left( \frac{z+\phi_0}{\beta} \right),0\right), \\ \nn   V(\vec x) =\mathrm{const.}
\end{eqnarray}
where $A>0$ and $\phi_0$ are integration constants. We can simplify it by Euclidean transformations and a shift of the potential to
\begin{eqnarray}\label{p1p2siAV}
 \nn \vec A(\vec x) =\left(-A \cos\left(\frac{z}{\beta} \right),-A \sin\left(\frac{z}{\beta} \right),0\right), \\
 \vec B(\vec x)=\left(\frac{A}{\beta} \cos\left( \frac{z}{\beta} \right), \frac{A}{\beta} \sin\left( \frac{z}{\beta} \right),0\right), \qquad  V(\vec x) =0.
\end{eqnarray}
The integral of motion $X_3$~\eqref{p1p2x3} reduces to
\begin{equation}
X_3=l_3+\beta p_3
\end{equation}
in the gauge chosen above. As before, this calculation is the same in both classical and quantum mechanics.

The classical equations of motion take the form
\begin{eqnarray}
\fl \nn \dot p_1(t)  = 0, \quad \dot p_2(t) = 0, \quad \dot p_3(t) = \frac{A}{\beta}  \left( -\sin\left(\frac{z(t)}{\beta}\right) p_1(t)+ \cos\left(\frac{z(t)}{\beta}\right) p_2(t)\right), \\
\fl \label{classeqmotp1p2} \dot x(t) = p_1(t)-A \cos\left(\frac{z(t)}{\beta}\right), \quad  \dot y(t) = p_2(t)-A \sin\left(\frac{z(t)}{\beta}\right), \quad \dot z(t)  = p_3(t)
\end{eqnarray}
and can be solved by quadratures. Namely, we express the conserved momenta in the polar form
\begin{equation}
p_1=p \cos\left(\frac{\phi_p}{\beta}\right),\qquad p_2=p\sin\left(\frac{\phi_p}{\beta}\right)
\end{equation}
where $p\geq 0$ and $\phi_p$ are constants, and find a second order equation for $z(t)$
\begin{equation}
\ddot z(t)=-\frac{A \, p}{\beta}\sin\left(\frac{z(t)-\phi_p}{\beta}\right).
\end{equation}
The order of this equation can be lowered, obtaining
\begin{equation}
\label{loweredODEforz} \frac{1}{2}\left(\dot z(t)\right)^2=A \, p \, \left( \cos\left(\frac{z(t)-\phi_p}{\beta}\right)+\kappa \right), \qquad \kappa \geq -1.
\end{equation}
($\kappa<-1$ is unphysical since then~\eqref{loweredODEforz} doesn't have real solutions.)
We substitute 
\begin{equation}
z(t) = \phi_p+\beta\arccos(\zeta(t))
\end{equation} and we find a separable first order ODE for $\zeta(t)$
\begin{equation}\label{ODEellints}
(\dot \zeta(t))^2 = -\frac{2A p}{\beta^2} (\zeta(t)-1) (\zeta(t)+1) (\zeta(t)+\kappa)
\end{equation}
solvable in terms of elliptic integrals. Explicitly, we change the independent variable
\begin{equation}
t = \frac{\beta}{\sqrt{2 A p}} \tau
\end{equation}
to have a simpler equation
\begin{equation}
(\dot \zeta(\tau))^2 = - (\zeta(\tau)-1) (\zeta(\tau)+1) (\zeta(\tau)+\kappa).
\end{equation}
The solution depends on the value of the integration constant $\kappa$, namely $1<\kappa<-1$ and $1\leq \kappa$ demonstrate  different behavior, and on the initial value for $\zeta(\tau)$.
For $\kappa>1$ we find a solution in the form
\begin{eqnarray}
\zeta(\tau)  = \frac{1-\kappa^2}{2\mathrm{sn}^2\left(\frac{1}{2} \sqrt{\kappa+1}(\tau-\tau_0), \sqrt{\frac{2}{\kappa+1}}\right)-\kappa-1}-\kappa.
\end{eqnarray}
For $-1<\kappa<1$ we find a solution in the form
\begin{eqnarray}
\zeta(\tau)  = \frac{2 (1-\kappa)}{2-(\kappa+1) \mathrm{sn}^2\left(\frac{1}{2} \sqrt{2} (\tau-\tau_0), \sqrt\frac{\kappa+1}{2}\right)}-1.
\end{eqnarray}

The equations of motion for $x(t),y(t)$ now reduce to quadratures~\eqref{classeqmotp1p2} in terms of the Jacobi elliptic function $\mathrm{sn}$. Solving them numerically we obtain the trajectories for our system. For $-1< \kappa< 1$ they are bounded in the plane perpendicular to $(p_1,p_2,0)$ and appear like a deformed helix whose axis is parallel to the vector $(p_1,p_2,0)$.
\begin{figure}[ht!]
\centering
\includegraphics[width=60mm]{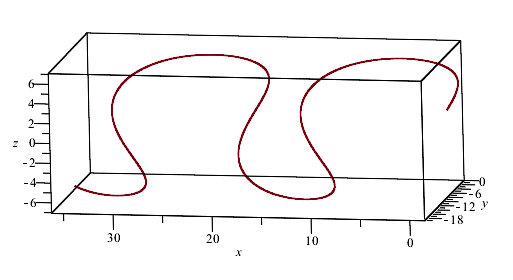}
\caption{Sample trajectory for $-1<\kappa<1$ (with $A=3,\beta=3,p_1=1,p_2=0, x(0)=0.08,y(0)=0.05,z(0)=0,\dot z(0)=3.2$)}
\label{figure1}
\end{figure}
For $1\leq \kappa$ they are no longer bounded in the $z$-direction and appear like a deformed helix whose axis is no longer parallel to the $xy$-plane.
\begin{figure}[ht!]
\centering
\includegraphics[width=60mm]{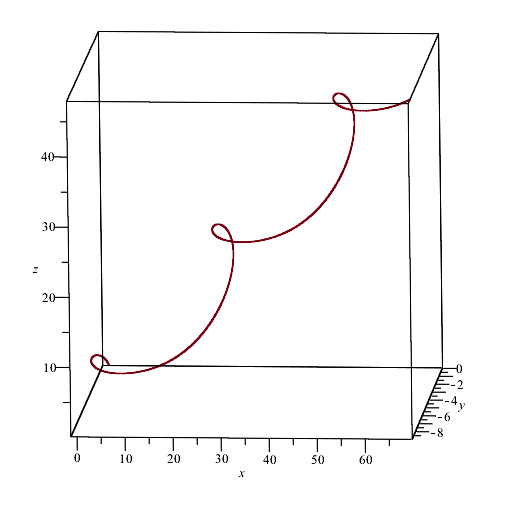}
\caption{Sample trajectory for $\kappa>1$ (with $A=3,\beta=3,p_1=1,p_2=0, x(0)=0.08,y(0)=0.05,z(0)=0.1,\dot z(0)=3.5$)}
\label{figure2}
\end{figure}
The value $\kappa=1$ appears to be a limiting case of the $\kappa>1$ situation.
\begin{figure}[ht!]
\centering
\includegraphics[width=60mm]{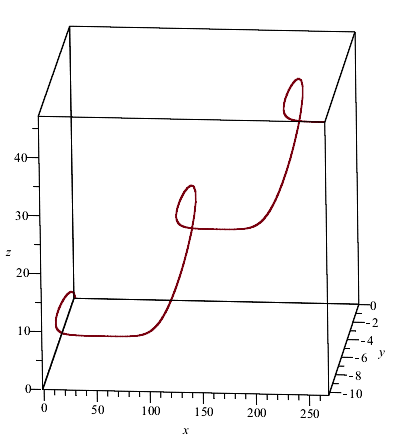}
\caption{Sample trajectory for $\kappa=1$ (with $A=3,\beta=3,p_1=1,p_2=0, x(0)=0.08,y(0)=0.05,z(0)=0,\dot z(0)=2\sqrt{3}$)}
\end{figure}

In the quantum case the stationary Schr\"odinger equation 
$$ \hat H \psi(\vec x) = E \psi(\vec x)$$
separates in Cartesian coordinates. We have
\begin{eqnarray}
\fl \nn \psi(\vec x) = \chi(z) \exp\left(\ii \frac{K}{\hbar} \cos(\phi_K)\, x\right) \exp\left(\ii \frac{K}{\hbar}\sin(\phi_k)\, y\right), \\
\fl X_1 \psi(\vec x) = K\cos(\phi_K) \psi(\vec x), \qquad X_2 \psi(\vec x) = K\sin(\phi_K) \psi(\vec x), \\
\nn \fl \hbar^2 \ddot \chi(z) = \left(-2 A K \cos\left(\frac{z}{\beta}-\phi_K\right)+A^2+K^2-2 E\right) \chi(z).
\end{eqnarray}
The separated equation for $\chi(z)$ is solved in terms of Mathieu sine and cosine functions $C(a,q,x)$ and $S(a,q,x)$, i.e. independent solutions of the Mathieu's differential equation 
$$\ddot y+(a-2 q \cos(2 x)) y(x) = 0,$$ as follows
\begin{eqnarray}\label{Mattsol}
\fl \chi(z)= C_1\, C\left(-4 \frac{\beta^2}{\hbar^2} (A^2+K^2-2 E), -4 \frac{\beta^2}{\hbar^2} A K, \frac{\phi_k}{2}-\frac{z}{2 \beta}\right)+ \\
\nn + C_2\, S\left(-4 \frac{\beta^2}{\hbar^2} (A^2+K^2-2 E), -4 \frac{\beta^2}{\hbar^2} A K, \frac{\phi_k}{2}-\frac{z}{2 \beta}\right).
\end{eqnarray}

This leads to the conjecture that the Hamiltonian system with the potentials~\eqref{p1p2siAV} is maximally superintegrable but an explicit calculation shows that that hypothetical further integral cannot be of order one or two in momenta. In the classical case an additional fifth integral can be found from the Hamiltonian flow~\eqref{classeqmotp1p2}. We use the method of characteristics to arrive at the equation
\begin{equation}
\frac{\mathrm{d}y }{p_2-A \sin\left(\frac{z}{\beta}\right)} = \frac{\mathrm{d}z }{\sqrt{2 A \cos(\frac{z}{\beta}) p_1+2 A \sin\left(\frac{z}{\beta}\right) p_2+u}}
\end{equation}
where $u=p_3^2-2 A \left(p_1 \cos(\frac{z}{\beta}))+p_2 \sin\left(\frac{z}{\beta}\right)\right)=2 H -X_1^2-X_2^2-A^2
$ is a constant of motion.
Its solution is expressed in terms of Jacobi elliptic functions whose arguments depend on the momenta $p_1,p_2$ and $p_3$. Hence is not polynomial in the momenta at all.
\end{itemize}

\section{Superintegrability for the integrable system with integrals $L_3,P_3$}
Let us perform a similar analysis for the case~\eqref{l3p3}  
$$ X_1=l_3^A+m_1(\vec x),\qquad X_2=p_3^A+m_2(\vec x).$$
Requiring that $X_1$ and $X_2$ are in involution we find the condition
\begin{equation}
x B_1+y B_2+x \partial_y m_2-y \partial_x m_2-\partial_z m_1=0.
\end{equation}
Equations~\eqref{1ordcond} reduce to
\begin{eqnarray}
\nn \partial_x m_1 = -x B_3 ,  \qquad \partial_y m_1 = -y B_3 ,  \qquad \partial_z m_1 = y B_2 +x B_1, \\
\partial_x m_2 = B_2 , \qquad  \partial_y m_2 = -B_1 , \qquad  \partial_z m_2 = 0.
\end{eqnarray} 
Solving these equations and their compatibility conditions we find that
\begin{eqnarray}
\nn m_1(\vec x)  = -F_2(R), \qquad  m_2(\vec x)  = F_1(R), \qquad  R  =\sqrt{x^2+y^2},\\ 
\label{l3p3ABM} \vec B(\vec x)  =\left( - F'_1 \frac{y}{R},  F'_1 \frac{x}{R},  \frac{1}{R} F'_2 \right),\\
\nn \vec A(\vec x)  = \left( -\frac{y}{R^2} F_2(R) , \frac{x}{R^2} F_2(R),  -F_1(R)\right),\qquad  V(\vec x)  =V(R).
\end{eqnarray}
Substituting~\eqref{l3p3ABM} into our form of the integrals~\eqref{l3p3} we find that in our choice of gauge we have in fact
\begin{equation}\label{l3p3final}
X_1=l_3,\qquad X_2=p_3,
\end{equation}
i.e. the first order integrals are again of direct geometric origin and there is no other possibility for them if we assume their form as in~\eqref{l3p3}.

The computation in the quantum case is essentially the same. We obtain the same structure of the potentials~\eqref{l3p3ABM} and the integrals
\begin{equation}\label{l3p3finalqm}
\hat X_1=\hat L_3,\qquad \hat X_2=\hat P_3.
\end{equation}
Thus the stationary Schr\"odinger equation separates in the polar coordinates
\begin{eqnarray}
 x  =R \cos{\phi},  \qquad y=R \sin{\phi}, \qquad  z = z
\end{eqnarray}
as follows
\begin{eqnarray}
\nn \psi(\vec x)  = &\exp(\ii m \phi) \exp(\ii k z) \rho(R),\\
\hbar^2 \ddot \rho(R) = & -\hbar^2 \frac{\dot \rho(R)}{R}+
\left(  F_1(R)-\hbar k\right)^2 \rho(R)+2 V(r) \rho(R)\\
& \nn  +\frac{1}{R^2}\left( F_2(R)+\hbar m  \right)^2 \rho(R)-2 E \rho(R).
\end{eqnarray}

Like in the previous section let us now turn our attention towards possible superintegrable Hamiltonians which are integrable by virtue of the integrals~\eqref{l3p3final}. We assume first an additional integral in the first order form
\begin{equation}
X_3= \sum_{i=1}^{2} \left(\gamma_3^i l_i+\beta_3^i p_i\right)+m_3(\vec x).
\end{equation}
As before, the compatibility of equations~\eqref{1ordcond} strongly restricts the possibilities. Namely,
\begin{itemize}
\item if $(\gamma_3^1)^2+(\gamma_3^2)^2\neq 0$ then the magnetic field $\vec B(\vec x)$ must vanish, i.e. this case is of no interest to us here;
\item if $\gamma_3^1=\gamma_3^2= 0$ then we arrive at the already discussed superintegrable case of the constant magnetic field and vanishing electric field.
\end{itemize}
Thus we have not found any nontrivial Hamiltonian with magnetic field superintegrable at the first order with the integrals~\eqref{l3p3final}.

Next we consider the same problem with the second order integral of the form~\eqref{classintUEA}. By subtracting a function of the known integrals and the Hamiltonian and using the relation $\vec p\cdot \vec l=0$ we  can a priori set to
zero the constants
\begin{equation}
\alpha_{11}, \alpha_{14}, \alpha_{33}, \alpha_{36}, \alpha_{66}.
\end{equation}
We substitute these into~\eqref{classintUEA} and consider equations~\eqref{2ordcond}, \eqref{1ordcond} and \eqref{0ordcond} and their compatibility conditions. We find after a tedious but straightforward calculation whose details we are not presenting here that for nonconstant functions $F_1$ and/or $F_2$ in~\eqref{l3p3ABM} no second order integral~\eqref{classintUEA} independent of the Hamiltonian, $X_1$ and  $X_2$ can be found. 

Thus we have to conclude that the system with the potentials and field strength~\eqref{l3p3ABM} is not first or second order minimally superintegrable for any nonconstant choice of the magnetic field $\vec B(\vec x)$ and the electrostatic potential $V(\vec x)$. The same result applies also to the quantum case where only the difference between equations~\eqref{0ordcond} and~\eqref{0ordcondquant} needs to be considered.

\section{Superintegrability for the integrable system with integrals $L_1,L_2,L_3$}\label{Secl1l2l3}
Let us now turn our attention to the case when we have three first order integrals of motion~\eqref{l1l2l3}. We cannot choose among them two in involution but we easily obtain a second order integral
\begin{equation}
(\vec X)^2=(X_1)^2+(X_2)^2+(X_3)^2
\end{equation}
which is in involution with all of them. Thus assuming that we have the integrals
$$ X_1=l_3^A+m_1(\vec x),\qquad X_2=l_1^A+m_2(\vec x)$$
we have immediately a minimally superintegrable system.

The compatibility of equations~\eqref{1ordcond} for the three integrals $X_1,X_2$ and $X_3=l_2^A+m_3(\vec x)$ leads directly to the following 1--parameter family of solutions for the magnetic field
\begin{equation}\label{l1l2l3B}
\vec B(\vec x)= g \frac{\vec x}{|\vec x|^3},
\end{equation}
i.e. the only possibility is a magnetic monopole of an arbitrary strength $g$. The vector potential $\vec A$ is always singular at least along a halfline connecting the origin to infinity. We can take e.g.
\begin{eqnarray}\label{l1l2l3A}
\vec A(\vec x) = \frac{g}{|\vec x| (x^2+y^2)} \left( y(z-|\vec x|), -x(z-|\vec x|), 0\right)
\end{eqnarray} 
which satisfies the Coulomb gauge condition $\nabla \vec A=0$.
The functions $m_j(\vec x)$ are obtained by integrating equations~\eqref{1ordcond} and up to an irrelevant additive constant read
\begin{equation}
m_j(\vec x) = g \frac{x_j}{|\vec x|}.
\end{equation}
From the condition~\eqref{0ordcond} we find that the electrostatic potential $V(\vec x)$ must be spherically symmetric,
\begin{equation}\label{l1l2l3V}
V(\vec x)=V(|\vec x|).
\end{equation}
Thus the classical Hamiltonian system~\eqref{classham} with the potentials and field strengths defined in~\eqref{l1l2l3B}, \eqref{l1l2l3A}, \eqref{l1l2l3V} is the only system which possesses the three first order integrals~\eqref{l1l2l3} and is minimally superintegrable due to the functionally independent integral $X_3$.  Explicitly, the integrals of motion in our choice of gauge~\eqref{l1l2l3A} read
\begin{eqnarray}
\nn X_1 =l_1+g \frac{x (|\vec x|-z)}{x^2+y^2}, \\
X_2 =l_2+g \frac{y (|\vec x|-z)} {x^2+y^2},\\
\nn X_3 =l_3+g,\\
\nn (\vec X)^2 =(l_1)^2+(l_2)^2+(l_3)^2+2 g (l_3+g) |\vec x| \frac{|\vec x|-z}{x^2+y^2}.
\end{eqnarray}
Since all the conditions considered are the same for both the classical and quantum situation, we have the same structure of minimally superintegrable system also at the quantum level. This rotational invariance of the magnetic monopole was already observed in~\cite{Peres}.\bigskip

Next, we shall look for an additional independent integral $X_4$ of the form~\eqref{classintUEA}, i.e. at most second order in momenta, which would make our system maximally superintegrable. That means looking at the conditions ~\eqref{HOconds}-\eqref{0ordcond} for the already determined magnetic field $\vec B$~\eqref{l1l2l3B} and restricted electrostatic potential $V(|\vec x|)$~\eqref{l1l2l3V} and establishing for which choices of $V(\vec x)$ an additional integral exists. We assume that a suitable polynomial combination of the Hamiltonian and the already known integrals $X_1,X_2,X_3$ was subtracted from $X_4$. Together with the relation 
$$\vec l \cdot \vec p=0$$ 
we can thus set to zero the following constants $\alpha_{ij}$ in the integral~\eqref{classintUEA}
$$ \alpha_{11}, \alpha_{14}, \alpha_{44}, \alpha_{45}, \alpha_{46}, \alpha_{55}, \alpha_{56}, \alpha_{66}.$$
The compatibility conditions for equations~\eqref{2ndderss} lead to the following values of the remaining constants $\alpha_{ij}$ in~\eqref{classintUEA}
\begin{eqnarray}
\nn \alpha_{12}  = 0,\qquad  \alpha_{13}  = 0,\qquad  \alpha_{22} = 0,\qquad  \alpha_{23}  = 0, \\
 \alpha_{24}  = -\alpha_{15},\qquad  \alpha_{25}  = 0,\qquad  \alpha_{33}  = 0,   \\
\nn \alpha_{34}  = -\alpha_{16},\qquad  \alpha_{35}  = -\alpha_{26},\qquad  \alpha_{36}  = 0  
\end{eqnarray}
leaving three yet undetermined constants $\alpha_{15},\alpha_{16},\alpha_{26}$.
Solving the conditions~\eqref{2ordcond} for $\vec s$ we find
\begin{eqnarray}
\nn s_1(\vec x)  =  g \left(\alpha_{15}\frac{y}{|\vec x|}+  \alpha_{16}\frac{z}{|\vec x|}\right),\\ 
\label{monopoleS} s_2(\vec x)  = g\left( \alpha_{26}\frac{z}{|\vec x|}- x \alpha_{15}\frac{x}{|\vec x|}\right),\\
\nn s_3(\vec x)  = -g\left( \alpha_{16}\frac{x}{|\vec x|}+ \alpha_{26}\frac{y}{|\vec x|}\right).
\end{eqnarray}
The compatibility of the conditions~\eqref{1ordcond}, e.g. $\partial_x(\partial_y m)=\partial_y(\partial_x m)$, requires that either 
\begin{itemize}
\item $\alpha_{15},\alpha_{16},\alpha_{26}$ are all equal to zero and thus also $\vec s=0$, i.e. there is no additional independent integral, or
\item the scalar potential must satisfy
\begin{equation}\label{monopoleCoulomb}
V(\vec x) = \frac{g^2}{2} \frac{1}{|\vec x|^2}-\frac{Q}{|\vec x|},
\end{equation}
i.e. the particle moves in the Coulomb potential modified by the $|\vec x|^{-2}$ term proportional to the strength of the magnetic monopole. Solving the remaining conditions~\eqref{1ordcond} we find that the scalar part of the integral $X_4$ now reads
\begin{equation}\label{monopoleCoulombm}
m(\vec x) = \frac{2 Q}{|\vec x|} \left( \alpha_{16} y-\alpha_{26} x-\alpha_{15} z \right).
\end{equation}
The condition~\eqref{0ordcond} is satisfied identically after plugging-in~\eqref{monopoleS} and~\eqref{monopoleCoulomb}.

The three constants $\alpha_{15},\alpha_{16},\alpha_{26}$ remain arbitrary and we have three additional integrals of the given form which are the components of the Laplace-Runge-Lenz vector modified by the presence of the magnetic monopole
\begin{equation}
R_j= \epsilon_{jkl} \left( p_k+A_k \right) X_l - Q \frac{x_j}{|\vec x|}, \qquad j=1,2,3.
\end{equation}
Of course, only one of them is functionally independent of the Hamiltonian and the integrals $X_1,X_2,X_3$.

The same conclusions apply also to the quantum case where the analysis is essentially the same, taking into account appropriate symmetrization, and the expressions \eqref{l1l2l3B}, \eqref{monopoleS}, \eqref{monopoleCoulomb} and~\eqref{monopoleCoulombm} can be taken over literally.

The fact that the system with the magnetic field~\eqref{l1l2l3B} and the modified Coulomb potential~\eqref{monopoleCoulomb} is maximally superintegrable has of course been known for long time (see e.g. \cite{McICis,LaMayVi1} and references therein). Here we have shown that under the restrictions imposed on the structure and order of the integrals there is no other maximally superintegrable case in this class. We notice that the restrictions imposed are more general than the ones under which a similar result was derived in \cite{HrBal}.

While it may be surprising that no modification of the isotropic harmonic oscillator arose in our calculation, we refer the reader to \cite{LaMayVi1,LaMayVi2} where it was demonstrated that it is maximally superintegrable but with fourth order integrals, not at most second, as considered here.
\end{itemize} 

\section{Conclusions}
In Section~\ref{ConIntMot} we derived the determining equations \eqref{HOconds}--\eqref{0ordcond} for the coefficients of a general second order integral of motion~\eqref{classint} and discussed their compatibility conditions. As in the case of a purely scalar potential~\cite{WinSmoUhlFr} the coefficients $h_j(\vec x), n_j(\vec x)$ satisfy equations~\eqref{HOconds} that do not depend on the potentials. These equations~\eqref{HOconds} are easy to solve and imply that the leading terms in the integral $X$ lie in the enveloping algebra of the Euclidean Lie algebra $\mathfrak{e}_3$. As opposed to the scalar case $(\vec A(\vec x)=0)$, first order terms in $X$ are not excluded (even and odd terms do not commute separately). Also, for $\vec A(\vec x)\neq 0$ the classical and quantum determining equations differ, see~\eqref{0ordcondquant}. We recall that for scalar particles the classical and quantum determining equations and hence also the integrals of motion and the superintegrable Hamiltonians differ only for integrals of order $N\geq 3$~\cite{PoWi,GraWi,Gravel,Mar1,Mar2}. 

In general, these determining equations and also their compatibility conditions are difficult to solve. Indeed, if the magnetic field $\vec B(\vec x)$ and the potential $V(\vec x)$ are not known the equations are nonlinear. They can be used in several manners. First of all, for the system to be second order integrable two such integrals must exist in addition to the Hamiltonian and they must commute. The leading order terms of such pairs of commuting integrals were classified into 11 conjugacy classes~\cite{MaSmoVaWin} under Euclidean transformations and this classification remains the same for $\vec B(\vec x)\neq 0$ though the nonleading terms are different. In the purely scalar case each class corresponds to the separation of variables in the Hamilton-Jacobi equation and in the Schr\"odinger equation, respectively.

This correspondence no longer holds in the presence of a magnetic field. However, for $\vec B(\vec x)\neq 0$ it is still possible to consider each class separately and this allows significant simplifications. The case of integrals of motion with leading terms of the form $P_1^2,P_2^2$ was studied in~\cite{Zhalij}.

It is actually easier to study superintegrable systems than integrable ones since the conditions on the potentials are more constraining. In this article we have determined all superintegrable systems with at least 2 first order integrals and at least four independent integrals altogether (including the Hamiltonian). We have found the following superintegrable systems:
\begin{enumerate}
\item A constant magnetic field and no electric field~\eqref{constBpots}. This system has 4 first order integrals~\eqref{constBints}, the Hamiltonian is a polynomial in these four. The system is maximally superintegrable but the fifth integral $X_5$~\eqref{constB5int} is a transcendental function of the momenta rather than a polynomial. In classical mechanics this is not a problem. In quantum mechanics there are difficulties with its interpretation. This system is well-known to be exactly solvable~\cite{LanLif,Newton} but to our knowledge its maximal superintegrability has not be noticed before. The Poisson algebra of integrals of motion is given in~\eqref{tabcomms}.
\item The system~\eqref{p1p2siAV} with a periodic magnetic field and zero electric one. There are 3 first order integrals $p_1,p_2$ and $l_3+\beta p_3$, the Hamiltonian $H$ and one nonpolynomial integral. The classical trajectories are given in terms of elliptic functions~\eqref{ODEellints} and examples are given on Figure~\ref{figure1} and Figure~\ref{figure2}. The Sch\"odinger equation is solved in terms of periodic Mathieu functions~\eqref{Mattsol}. The energy spectrum is continuous.
\item The magnetic monopole with the magnetic field~\eqref{l1l2l3B} and the scalar potential~\eqref{monopoleCoulomb}. Its second order maximal superintegrability is well-known~\cite{McICis}. We have shown that it is the only second order spherically symmetric maximally superintegrable system in $E_3$ with nonvanishing magnetic field. A fourth order superintegrable system is also known~\cite{LaMayVi1,LaMayVi2}. 
\end{enumerate}
All maximally superintegrable systems presented in this article are exactly solvable, both in classical and quantum mechanics. To show their superintegrability it was necessary to consider nonpolynomial and nonrational integrals of motion. In a completely different context nonpolynomial integrals arise also for the purely scalar potentials, see e.g.~\cite{Hie1,Hie2,Feh1,Feh2}.

In \cite{BeWin} the structure of the gauge--invariant integrable and superintegrable systems involving vector potentials was considered in two spatial dimensions. Among other results it was shown there that under the assumption that the integrals are at most second order in momenta every superintegrable system in dimension 2 has constant magnetic field. However, as we have seen in Section~\ref{Secl1l2l3} in three spatial dimensions the second order maximal superintegrability does not imply constant magnetic field.

\section*{Acknowledgments}
The research of A. M. was supported by the European social fund within the framework of realizing the project "Support of inter-sectoral mobility and quality enhancement of research teams at Czech Technical University in Prague", CZ.1.07/2.3.00/30.0034.

L. \v{S}. was supported by the Grant Agency of the Czech Technical University in Prague, grant No. SGS 13/217/OHK4/3T/14 and by the Czech Ministry of Education, RVO68407700. 

P. W. was partially supported by a research grant from NSERC of Canada. He thanks the European Union Research Executive Agency for the award of a Marie Curie International Incoming Award Fellowship that made his stay at the University Roma Tre possible. He thanks the Department of Mathematics and Physics of the University Roma Tre and specially Professor D. Levi for hospitality.


\end{document}